\documentclass[twocolumn,prb,aps,longbibliography,superscriptaddress,floatfix,
tightenlines]{revtex4-1}
\usepackage{amsmath}
\usepackage{amssymb}
\usepackage{mathrsfs}
\usepackage{amsfonts}
\usepackage{epsfig}
\usepackage{bm}% bold math
\usepackage{times}
\usepackage[T1]{fontenc}
\usepackage[dvips]{color}
\usepackage[colorlinks,bookmarks=false,citecolor=blue,linkcolor=magenta,urlcolor=blue]{hyperref}

\begin{document}

%\title{Observation of information backflow from controllable non-Markovian environments in diamond}
\title{Observing information backflow from controllable non-Markovian multi-channels in diamond}
\author{Ya-Nan Lu}
\thanks{These authors contributed equally to this work.}
\affiliation{Institute of Physics, Chinese Academy of Sciences, Beijing 100190, China}
\affiliation{School of Physical Sciences, University of Chinese Academy of Sciences, Beijing 100049, China}

\author{Yu-Ran Zhang}
\thanks{These authors contributed equally to this work.}
\affiliation{Theoretical Quantum Physics Laboratory, RIKEN Cluster for Pioneering Research,
Wako-shi, Saitama 351-0198, Japan}
%\affiliation{Beijing Computational Science Research centre, Beijing 100193, China}

%\affiliation{Beijing National Laboratory for Condensed Matter Physics, Institute of Physics, Chinese Academy of Sciences, Beijing 100190, China}
\author{Gang-Qin Liu}
\email{gqliu@iphy.ac.cn}
\affiliation{Institute of Physics, Chinese Academy of Sciences, Beijing 100190, China}
\affiliation{Songshan Lake Materials Laboratory, Dongguan, Guangdong 523808, China}
%\affiliation{CAS center of Excellence in Topological Quantum Computation, Beijing 100190, China}

%\author{J. Q. You}
%%\email{jqyou@zju.edu.cn}
%\affiliation{Department of Physics, Zhejiang University, Hangzhou 310027, China}
%\affiliation{Beijing Computational Science Research centre, Beijing 100193, China}

\author{Franco Nori}
%\email{fnori@riken.jp}
\affiliation{Theoretical Quantum Physics Laboratory, RIKEN Cluster for Pioneering Research, Wako-shi,
Saitama 351-0198, Japan}
\affiliation{Physics Department, University of Michigan, Ann Arbor, Michigan 48109-1040, USA}

\author{Heng Fan}
\email{hfan@iphy.ac.cn}
\affiliation{Institute of Physics, Chinese Academy of Sciences, Beijing 100190, China}
\affiliation{Songshan Lake Materials Laboratory, Dongguan, Guangdong 523808, China}
\affiliation{CAS Centre of Excellence in Topological Quantum Computation, Beijing 100190, China}

\author{Xin-Yu Pan}
\email{xypan@aphy.iphy.ac.cn}
\affiliation{Institute of Physics, Chinese Academy of Sciences, Beijing 100190, China}
\affiliation{Songshan Lake Materials Laboratory, Dongguan, Guangdong 523808, China}
\affiliation{CAS Centre of Excellence in Topological Quantum Computation, Beijing 100190, China}

%\pacs{}

%\begin{abstract}
%
%%Since quantum Fisher information is a witness of multiparite entanglement useful for quantum metrology and
%
%
%
% %
%%not possible and since a complete microscopic description or
%%control of the environmental degrees of freedom is not feasible or only partially so.
%%
%%Quantum Fisher information flow, the dynamical evolution of quantum Fisher information (QFI), plays a critical role in non-Markovian characterization and quantum metrology.  Here, with the nitrogen-vacancy center in diamond as open system, we experimentally establish the transition from Markovian to non-Markovian process from the perspective of quantum Fisher information flow. Furthermore, by regarding the nitrogen nuclear spin and the surrounding strongly coupled ${}^{13}$C nuclear spin as different dissipative channels, we demonstrate that the total QFI flow can be decomposed into split contributions from individual channel. These results provide experimental insight into the relations between QFI flow and non-Markovianity in open system dynamics and pave the way for application in quantum parameter estimation.
%\end{abstract}

\maketitle
%\section{Introduction}
{\bf
\noindent Any realistic quantum system is inevitably subject to an external environment. This environment makes the
open-system dynamics significant for many quantum technologies, such as entangled-state
engineering \cite{Bradley2019,Omran2019,Song2019}, quantum simulation \cite{Georgescu2014}, and quantum sensing \cite{Degen2017}.
The information flow of a system to its environment usually induces a Markovian process, while the backflow
of information from the environment exhibits non-Markovianity \cite{Breuer2007}.
The practical environment, usually consisting of a large number of degrees of freedom, is hard to
control, despite some attempts on controllable transitions from Markovian to non-Markovian dynamics \cite{Myatt2000,Liu2011,Gessner2013,Haase2018,Andersson2019,Wu2019}.
Here, we experimentally demonstrate the engineering of multiple dissipative channels by controlling the adjacent nuclear spins of a nitrogen-vacancy centre in diamond.
With a controllable non-Markovian dynamics of the open system, we observe that the quantum Fisher information
flows \cite{Lu2010} to and from the environment using different noisy channels.
%Through the quantum Fisher information flows , we observe the controllable.
Our work contributes to the developments of
both noisy quantum metrology \cite{Giovannetti2011,Chin2012,Liu2016a} and quantum open systems  from the viewpoints
of metrologically useful entanglement.}

The unavoidable interaction of a quantum open system with its environment leads to the
dissipation of quantum coherence and correlations, making its dynamical behaviour %of an open system
a key role in many quantum technologies.
 According to the orientation of the information flow between the system and the environment,
 the time evolution of the open quantum system can be classified into either a Markovian or a non-Markovian one.
A Markovian process assumes memoryless
dynamics of a quantum open system, described by a dynamical semigroup
with a time-independent Lindblad generator \cite{Breuer2016}. However, %in many cases
in the presence of memory effects, e.g., for a strong system-environment coupling, the Markovian approximation
fails, and the non-Markovian process, deviating from a dynamical semigroup, allows for a revival of
quantum features.

\begin{figure}[b]
	\includegraphics[width=0.47\textwidth]{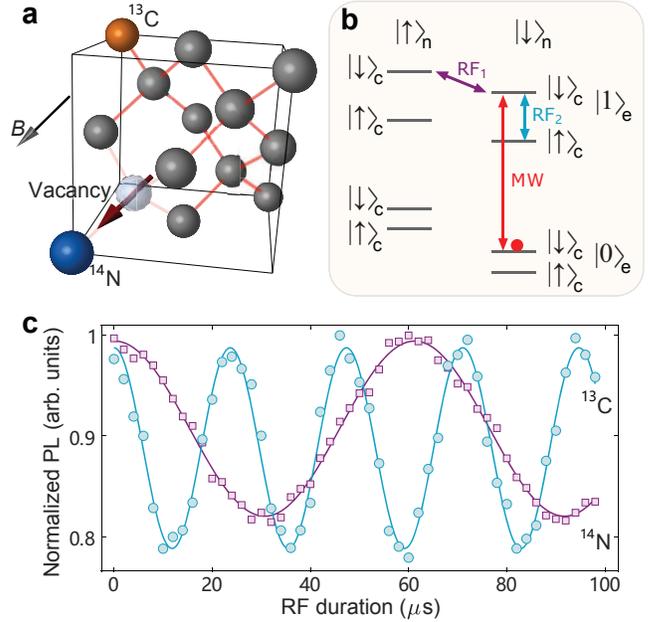}	
	\caption{{\bf Coherent manipulation of multiple spins in diamond.} (\textbf{a}) The nitrogen-vacancy (NV) centre, its  host ${}^{14}$N nuclear spin, and a nearby ${}^{13}$C nuclear spin form a three-qubit system.
(\textbf{b}) Energy levels of the three-qubit system.
At the excited-state level anti-crossing (ESLAC), the three spin qubits can be polarised by a short laser pulse and manipulated with resonant radio-frequency (RF) pulses (13.284~MHz for the ${}^{13}$C nuclear spin and 2.929~MHz for the host  ${}^{14}$N nuclear spin).
(\textbf{c}) Rabi oscillations of $^{14}$N and $^{13}$C nuclear spins under an external magnetic field of $B=482$~Gauss along the
quantisation axis of the NV centre (NV electron spin is at the $m_{s}=-1$ state).
}
	\label{fig0}
\end{figure}

\begin{figure*}[ht]
	\includegraphics[width=0.97\textwidth]{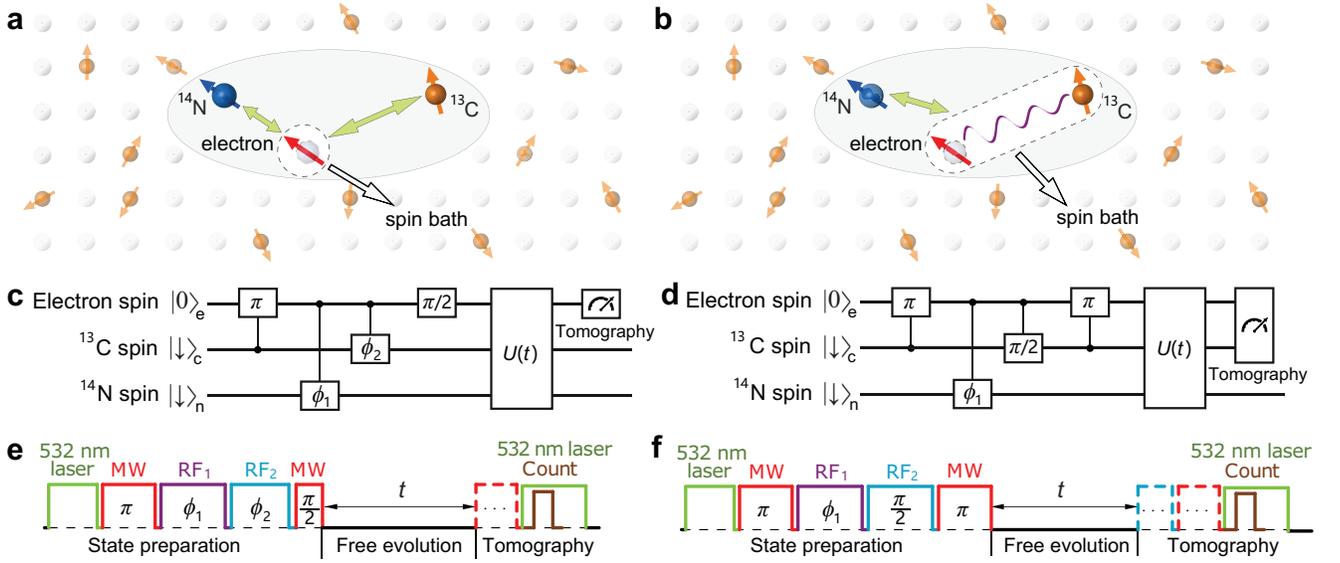}	
\caption{{\bf  Physical coding and experimental procedures.} The NV electron spin and two strongly coupled nuclear spins play the roles of the open system and controlled dissipative channels, respectively, while the other (weakly coupled) nuclear spins form an uncontrolled dissipative channel. The quantum Fisher information (QFI) is used to characterise the quantum coherence and metrologically useful entanglement of the open system. (\textbf{a}) In the first experiment, the NV electron spin is the open quantum system. The quantum coherence of the system is subject to two controllable dissipative channels formed by the ${}^{14}$N and $^{13}$C nuclear spins.  (\textbf{b}) In the second experiment, the electron spin and  the $^{13}$C nuclear spin form the open system, and the entanglement between the two spins is subject to a controllable dissipative channel formed by the ${}^{14}$N nuclear spin.	(\textbf{c},\textbf{d}) Quantum circuits and (\textbf{e},\textbf{f}) pulse sequences for the experiments.
The states of the controllable channel and the open system are prepared in sequence; and state tomography of the open system, which is used to calculate the QFI and the QFI flows, is carried out after different evolution durations.
%For this magnetic field, a level anti-crossing in the excited-state (ESLAC) renders simultaneous polarization of electron spin, nitrogen and 13C nuclear spin by optical pumping. Rabi oscillations of 13C and 14N nuclear spin are respectively manipulated by a 2.929 MHz and 13.284 MHz radio-frequency (RF) signal.
%(\textbf{c}) (\textbf{d})
}
	\label{fig1}
\end{figure*}

Owing to the memory effects and the ability of recovering quantum features, non-Markovian
quantum dynamics \cite{Zhang2012} opens a new perspective for applications in quantum metrology.
Quantum metrology \cite{Giovannetti2011,Pezze2018}, an emerging quantum technology, aims to use quantum resources to yield a higher
 precision of statistical errors in estimating parameters compared to classical approaches. %Quantum metrology behaves very different
However, the useful quantum coherence and multipartite entanglement for sub-shot-noise sensitivity,
quantified by the quantum Fisher information (QFI) \cite{BRAUNSTEIN1994,Ma2011}, are very fragile to decoherence.
Moreover, quantum metrological studies can be very
%Moreover, the behaviours of quantum metrological strategies are quite
different when being subject to either Markovian or non-Markovian noises \cite{Chin2012}.
Thus, it is important to establish an approach for characterising the
non-Markovianity of the open-system dynamics by using the QFI flow between the quantum
system and its environment \cite{Lu2010}.
The usual measures of non-Markovianity include: bipartite entanglement \cite{Rivas2010}, trace distance \cite{Breuer2009}, and
temporal steering \cite{Chen2016a}. However, the metrological approach based on QFI also works on the
information subflows through different dissipative channels for a class of time-local
master equations \cite{Lu2010}.
%Different from the usual or rigid measures of non-Markovianity,
% including bipartite entanglement \cite{Rivas2010}, trace distance  \cite{Breuer2009}, and
%temporal steering \cite{Chen2016a}, this metrological information approach, concerned with
% high-resolution and high-sensitivity measurements of physical parameters, also works on
%the information subflows according to different dissipative channels for a class of time-local
%master equations \cite{Lu2010}.

%Furthermore, it was proved that QFI flow as a sufficient for non-Markovian

%, quantum phase transitions, quantum information processing and dissipative quantum computation. However, there is no precise and corresponding definition in quantum non-Markovianity as classical non-Markovianity due to unique projective measurement postulate in quantum mechanism. More recently, several definitions of quantum non-Markovianity and methods for quantitative measurement of the degree for quantum memory effects have been introduced and developed in theory. Quantum non-Markovianity has been experimentally observed and witnessed by trace distance and entanglement in photonic systems, trapped-ion systems, nuclear magnetic resonance and nitrogen-vacancy center in diamond. The nitrogen-vacancy () in diamond is a well-suited experimental platform to further research into remaining problems in quantum non-Markovianity due to the real and fully controlled environment noise.

In our experiments, the open system is provided by a nitrogen-vacancy (NV) centre
electron spin, its host $^{14}$N nuclear spin, and a proximal $^{13}$C nuclear spin in diamond
(see Fig.~1a, and details of the sample are presented in Methods).
 The NV centre is a spin-1 system with a zero-field splitting of
$\Delta\simeq2.87$~GHz between $|0\rangle_{\textrm{e}}$ and $|\pm1\rangle_{\textrm{e}}$
of the ground spin triplet.
An external magnetic field along the NV symmetry axis is applied to degenerate
$|\pm1\rangle_{\textrm{e}}$ states. Here, the first qubit is encoded on the $|0\rangle_{\textrm{e}}$
and $|-1\rangle_{\textrm{e}}$ (hereafter labelled as $|1\rangle_{\textrm{e}}$) subspace of
the NV electron spin. The second and third qubits are encoded on the $\mid\uparrow\rangle_{\textrm{n,c}}$
and $\mid\downarrow\rangle_{\textrm{n,c}}$ states of the host $^{14}$N nuclear spin
and the nearby $^{13}$C nuclear spin, respectively (see Fig.~1b for the energy levels). Applying the secular approximation and ignoring the weak nuclear-nuclear
dipolar interactions, the effective interaction Hamiltonian of the three-qubit system and the spin bath can be written as \cite{Doherty2013}
(we set $\hbar=1$)
\begin{equation}
\hat{\mathcal{H}}_I=A_{\textrm{n}}^{\|}\hat{S}_{\textrm{e}}^z\hat{I}_{\textrm{n}}^z+A_{\textrm{c}}^{\|}\hat{S}_{\textrm{e}}^z\hat{I}_{\textrm{c}}^z+\hat{H}_{\textrm{R}},
\end{equation}
where $\hat{H}_\textrm{R}$ is the interaction Hamiltonian between the electron qubit and the spin bath,
$A_{\textrm{n}}^{\|}\simeq-2.16$~MHz and $A_{\textrm{c}}^{\|}\simeq12.8$~MHz denote
the hyperfine coupling strengths between the electron spin and nearby nuclear spins, respectively.
%given in optically detected magnetic resonance (ODMR) \cite{Gruber1997} spectra of the electron spin.
Under an external magnetic field of $B_z=482$~Gauss, the electron spin and two nuclear spins
can be simultaneously polarised by a short laser pumping due to level anti-crossing in the excited
state (ESLAC) \cite{Jacques2009}. In Fig.~1c, we show Rabi oscillations of the $^{13}$C and $^{14}$N
nuclear spins, driven by 13.284~MHz and 2.929~MHz radio-frequency (RF)
pulses, respectively.
We attribute the smooth and damping-free oscillations to
the superb coherence of the nuclear qubits and the well-controlled
driving pulses.
%The experiments
%are performed .

%{\color{red}[system descriptions, e.g., ODMR FID MW RF. effective
%Hamiltonian $\mathcal{H}=H_{\textrm{S}}+H_{\textrm{R}}$] [first qubit is encoded
%on the $|0\rangle_{\textrm{e}}$ and $|-1\rangle_{\textrm{e}}$ (here in after labelled as
%$|1\rangle_{\textrm{e}}$ subspace.]} (Our experiments can be divided into two parts....)

\begin{figure*}[ht]
	\includegraphics[width=0.97\textwidth]{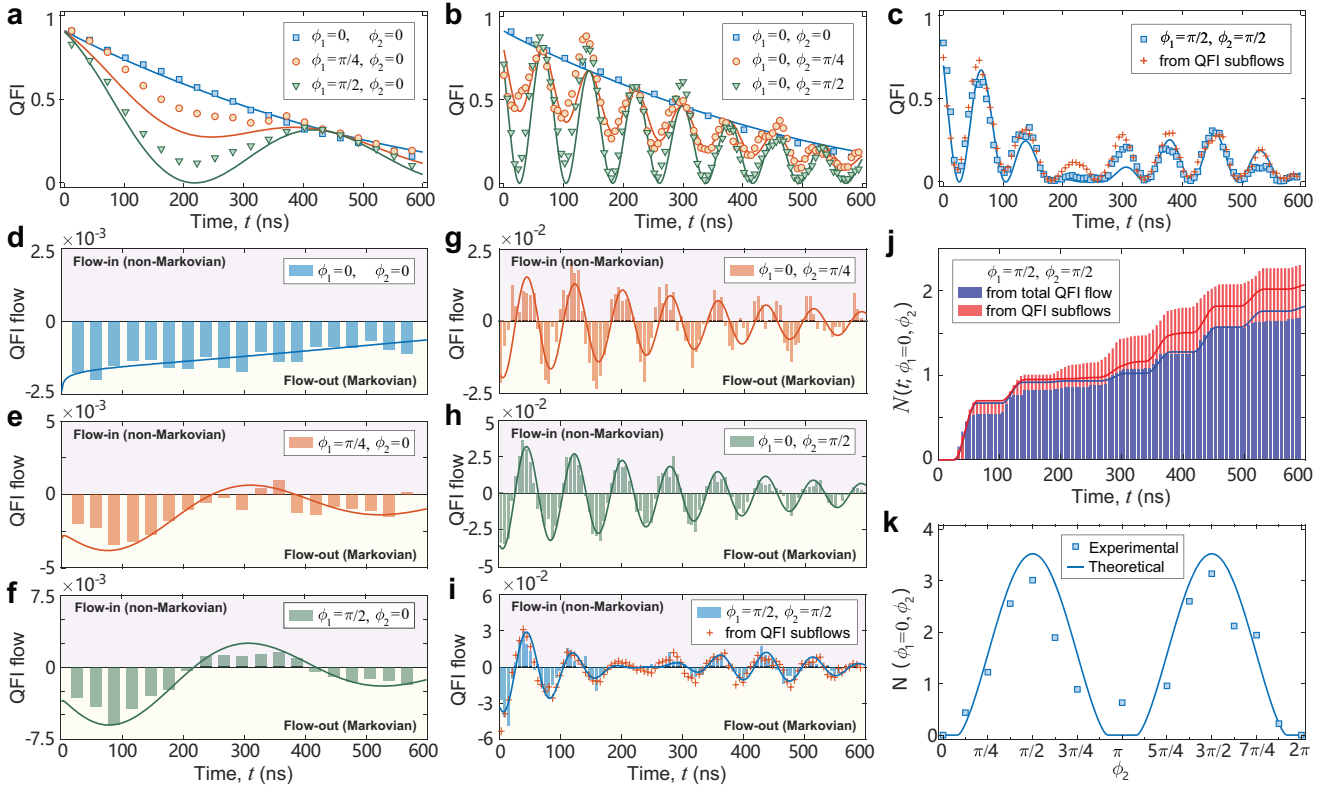}	
	\caption{{\bf Quantum Fisher information (QFI) flows of the electron qubit in a controllable non-Markovian environment.}
	 Time evolution of the QFI with (\textbf{a}) the dissipative channel of ${}^{14}$N open
	($\phi_1=\pi/4,\pi/2$) and the one of ${}^{13}$C closed ($\phi_2=0$);
	(\textbf{b}) the dissipative channel of ${}^{14}$N closed ($\phi_1=0$) and the one of ${}^{13}$C open ($\phi_2=\pi/4,\pi/2$);
	(\textbf{c}) the dissipative channels of both nuclear spins open ($\phi_1=\phi_2=\pi/2$) compared with
	the one calculated using the results of single channels given in (\textbf{a},\textbf{b}).
	QFI flows for the controllable dissipative channels with parameters:
	(\textbf{d}) $\phi_1=\phi_2=0$; (\textbf{e}) $\phi_1=\pi/4$, $\phi_2=0$; (\textbf{f}) $\phi_1=\pi/2$, $\phi_2=0$;
	(\textbf{g}) $\phi_1=0$, $\phi_2=\pi/4$; (\textbf{h}) $\phi_1=0$, $\phi_2=\pi/2$.
	(\textbf{i}) When the dissipative channels are adjusted with parameters $\phi_1=\phi_2=\pi/2$, the total QFI flow
	is compared with the sum of subflows calculated by results in (\textbf{d},\textbf{f},\textbf{h}).
	(\textbf{j}) The measure of non-Markovianity from the total positive QFI flow in (\textbf{i})
	compared with the measure  from
	the QFI subflows with respect to different dissipative channels [$N(t,\frac{\pi}{2},\frac{\pi}{2})$ in Eq.~(\ref{measure})],
	versus the evolution time.
	(\textbf{k}) The long-time non-Markovianity measure $\mathcal{N}(0,\phi_2)$ for different parameters $\phi_2$ of
	the $^{13}$C dissipative channel, when the $^{14}$N channel is closed ($\phi_1=0$).
	The solid curves are for the numerical simulations using experimental parameters.
}
	\label{fig2}
\end{figure*}

In our first experiment, we employ the electron qubit as a noise sensor,
whose dynamical behaviour is modulated by initialising the state of the
host $^{14}$N spin and the proximal ${}^{13}$C spin (see Fig.~2a).
Since the interactions between nuclear spins can be ignored in the time scale of our experiments,
these two controllable nuclear spins can be regarded as the regulators
 of two independent dissipative channels, and other weakly coupled nuclear spins act as another uncontrollable dissipative channel
(see Methods for details).
The quantum circuits and  pulse sequences of the first experiment
are shown in Figs.~2c and 2e, respectively.
By applying a 3~$\mu$s  laser pulse (532~nm),
these three qubits are polarised to an initial state
$|\Psi_{\textrm{i}}\rangle=|0\rangle_{\textrm{e}}\otimes\mid\downarrow\rangle_{\textrm{n}}\otimes\mid\downarrow\rangle_{\textrm{c}}$.
Then, by applying the MW, RF$_{1}$ and RF$_{2}$ pulses as shown in Fig.~2e, the system is prepared in
$|\Psi(0)\rangle=|+\rangle_{\textrm{e}}\otimes|\psi(\phi_1)\rangle_{\textrm{n}}\otimes|\psi(\phi_2)\rangle_{\textrm{c}}$, where
$|+\rangle_{\textrm{e}}\equiv(|0\rangle_{\textrm{e}}+|1\rangle_{\textrm{e}})/\sqrt{2}$, and
$|\psi(\phi)\rangle_{\textrm{n,c}}\equiv{\cos\frac{\phi}{2}\mid\uparrow\rangle_{\textrm{n,c}}}+\sin\frac{\phi}{2}\mid\downarrow\rangle_{\textrm{n,c}}$.
%Assumed that the prepared state of the system
%and the spin bath are separable as $\varrho_0=|\Psi(0)\rangle\langle\Psi(0)|\otimes\rho_{\textrm{R}}$,
The time evolution of the electron qubit can be described by the partial trace after
the unitary time evolution of the total Hamiltonian $\hat{U}(t)=\exp(-i\hat{\mathcal{H}}_It)$ as
$\rho_{\textrm{e}}(t)=\textrm{Tr}_{\textrm{ncR}}[\hat{U}(t)\varrho_0\hat{U}^\dag(t)]$,
where $\textrm{Tr}_{\textrm{ncR}}[\cdots]$ denotes the partial trace over the host $^{14}$N qubit,
the 12.8~MHz $^{13}$C qubit and the spin bath degrees of freedom. Given the generator $\hat{S}^{z}_{\textrm{e}}=\hat{\sigma}^z_{\textrm{e}}/2$, the QFI of the electron qubit can be written as
\begin{equation}
\mathcal{Q}(t;\phi_1,\phi_2)=r^2(t)-s_z^2(t)\simeq \mathcal{Q}_{\textrm{n}}(t;\phi_1)\mathcal{Q}_{\textrm{c}}(t;\phi_2)\mathcal{Q}_{\textrm{R}}(t),\label{eq3}
\end{equation}
where $r\equiv({s_x^2+s_y^2+s_z^2})^{\frac{1}{2}}$ is the length of the Bloch vector $\bm{r}=[s_{x}$, $s_{y}$, $s_{z}]$;
$\mathcal{Q}_{\textrm{n}}(t)=1-\sin^2\phi_1\sin^2({A^{\|}_{n}t}/2+\varphi_1/2)$,
$\mathcal{Q}_{\textrm{c}}(t)=1-\sin^2\phi_2\sin^2({A^{\|}_{c}t}/2+\varphi_2/2)$, and
$\mathcal{Q}_{\textrm{R}}(t)$
%=\exp[-(t/T_2^*)^\alpha]$, {\color{red}with $\alpha=1.1$ and $T_{2}^*\simeq362$~ns,}
are the QFI of the electron qubit only subject to the $^{14}$N, $^{13}$C, and the spin bath dissipative channels,
respectively.
%{\color{red}
	Details on %the form of
	$\mathcal{Q}_{\textrm{R}}(t)$
%the phases ($\varphi_1$ and $\phi_2$)
can be found in Methods.

%Supplementary Information. Note that the spin bath term follows a non-Gaussian decay ($\alpha\neq2$),
%indicating that the state distribution of the spin bath deviates from the normal thermal distribution \cite{Haase2018,Liu2012}.
%This case is reasonable when nearby nuclear spins are partially polarized at ESLAC \cite{Smeltzer2011}.}

%Here, the weak dipole-dipole interactions between nuclear spins are assumed to be neglected for the
%hold of the second equality in Eq.~(\ref{eq3}).

The QFI flow, defined as the rate of change of the QFI,
$\mathcal{I}(t)\equiv \partial_t \mathcal{Q}(t;\phi_1,\phi_2)$,
can be explicitly written as a sum of QFI subflows with respect to different dissipative channels \cite{Lu2010}
\begin{equation}
\mathcal{I}(t)=\mathcal{I}_{\textrm{n}}(t)+\mathcal{I}_{\textrm{c}}(t)+\mathcal{I}_{\textrm{R}}(t),
\end{equation}
where $\mathcal{I}_i(t)\equiv\mathcal{Q}(\partial_t\ln \mathcal{Q}_i)$, with $i=\textrm{n, c, R}$,
and each QFI subflow corresponds to not only the individual separable dissipative channel but
also all channels \cite{Lu2010}. Moreover, the inward QFI subflow ($\mathcal{I}_i>0$), resulting
from the temporary appearance of a
negative decay rate \cite{Rivas2014} of the time-local Lindblad master equation \cite{Breuer2016}, is
an essential feature of non-Markovian behaviours. Furthermore, we focus on
the sum of  time integrals of all inward QFI subflows
\begin{equation}
{N}(t,\phi_1,\phi_2)\equiv\sum_{i=\textrm{m,c,R}}\int_{0}^t\!\!\!dt\;\frac{|\mathcal{I}_i(\tau)|+\mathcal{I}_i(\tau)}{2},
\label{measure}
\end{equation}
as a measure of non-Markovianity, and the long-time measure is defined as
$\mathcal{N}(\phi_1,\phi_2)\equiv N(t\rightarrow\infty,\phi_1,\phi_2)$. %is defined as
%a measure of non-Markovianity.
Different from the time integral of the
total QFI flow, the measure of non-Markovianity in Eq.~(\ref{measure}) considers the inward
subflow from each dissipative channel and can dig out the non-Markovianity even when
the total QFI flow is negative.
%dynamics behaves Markovian.

%Taking the subflows into consideraction,
%the advantage of the measure in  over the time integral of the total QFI flow,

In our experiments, the dissipative channels of the $^{14}$N and $^{13}$C qubits can be fully controlled
by tuning the durations of the RF$_{1}$ and RF$_{2}$ pulses, i.e., to adjust $\phi_{1}$ and $\phi_2$.
%By tuning $\phi_1=0$ ($\phi_2=0$), the dissipative channel of $^{14}$N ($^{13}$C) nuclear spin is
%closed.
When both channels are turned off, $\phi_1=\phi_2=0$ (see Figs.~3a and 3b for the QFI),
the dynamics of the NV electron spin is only affected by the spin bath and behaves Markovian with the QFI
flowing out (see Fig.~3d). For $\phi_{1}=\pi/4,\pi/2$, and $\phi_{2}=0$, the channel of the $^{14}$N
nuclear spin is open, and the revival of the QFI is shown in Fig.~3a, while the positive QFI flows are observed
in Figs.~3e and 3f for witnessing non-Markovian dynamics. For $\phi_{1}=0$, and $\phi_{2}=\pi/4,\pi/2$,
the revival of the QFI and the positive QFI flows, subject to the channels of the $^{13}$C nuclear spin
and the spin bath, are plotted in Figs.~3b, 3g, and 3h.
For fifteen experimental instances of $^{13}$C qubit's parameter $\phi_2$, the measured $\mathcal{N}(0,\phi_2)$
is compared with the numerical simulation in Fig.~3k.

We furthermore characterise the system behaviour when both controllable dissipative channels are open.
With $\phi_1=\phi_2=\pi/2$, the time evolution of the QFI of the electron
qubit and its QFI flow, compared with the ones obtained from the sum of subflows (red cross), are
shown in Figs.~3c and 3i. In Fig.~3j, the measure of non-Markovianity $N(t,\frac{\pi}{2},\frac{\pi}{2})$
from the QFI subflows, $\mathcal{I}_{\textrm{n,c,R}}$, with respect to different dissipative channels (red bar) is compared with the
one from the total QFI flow, $\mathcal{I}$ (blue bar). We clearly observe that the measure in terms of QFI subflows
 quantify more non-Markovianity than the total QFI flow, when the system is subject to
multiple dissipative channels.

\begin{figure}[t]
	\includegraphics[width=0.42\textwidth]{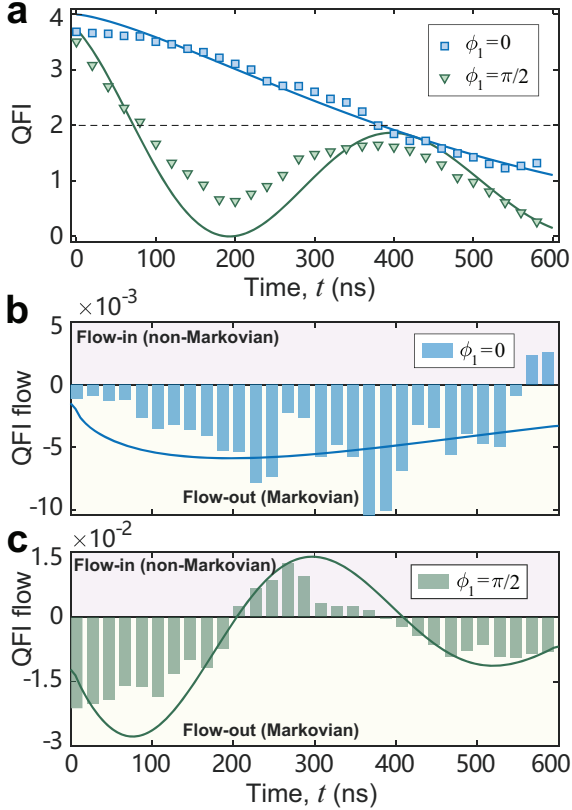}	
	\caption{{\bf Quantum Fisher information (QFI) flows of the two-qubit maximally entangled state
	in a controllable non-Markovian environment.}
	 (\textbf{a}) Time evolutions of the QFI of the maximally entangled state of the
	 NV electron qubit and the 12.8~MHz
	 $^{13}$C qubit with the controllable dissipative channel of the ${}^{14}$N qubit.
	QFI flows of the maximally entangled state
	 with the controllable dissipative channel of ${}^{14}$N (\textbf{b}) closed ($\phi_1=0$) and (\textbf{c}) open ($\phi_1=\pi/2$).
	The solid curves are for the numerical simulations using experimental parameters.
}
	\label{fig4}
\end{figure}

In the second experiment, we consider the open system, consisting of the electron qubit
and the proximal ${}^{13}$C qubit, which is subject to the controllable noisy channel of the host $^{14}$N
qubit and the dissipative channel of the spin bath (see Fig.~2b). Figures~2d and 2f show the quantum
circuit and  pulse sequences of the second experiment.
Starting with the state
$|\Psi_{\textrm{i}}\rangle=|0\rangle_{\textrm{e}}\otimes\mid\downarrow\rangle_{\textrm{n}}\otimes\mid\downarrow\rangle_{\textrm{c}}$,
%{\color{red}with a 5~$\mu$s 532nm laser pulse},
the system is prepared in
$|\Psi'(0)\rangle=(|0\rangle_{\textrm{e}}\otimes\mid\downarrow\rangle_{\textrm{c}}+|1\rangle_{\textrm{e}}\otimes\mid\uparrow\rangle_{\textrm{c}})\otimes|\psi(\phi_1)\rangle_{\textrm{n}}$,
with the pulse sequences shown in Figs.~2d and 2f.
The electron qubit and  ${}^{13}$C nuclear qubit are maximally entangled at this stage.
Similarly, assuming $\varrho_0'=|\Psi'(0)\rangle\langle\Psi'(0)|\otimes\rho_{\textrm{R}}$, the time
evolution of the electron qubit and the ${}^{13}$C qubit is described as
$\rho'_{\textrm{ec}}(t)=\textrm{Tr}_{\textrm{nR}}[\hat{U}(t)\varrho_0'\hat{U}^\dag(t)]$,
where $\textrm{Tr}_{\textrm{nR}}[\cdots]$ denotes the partial trace over the $^{14}$N spin
and the spin bath degrees of freedom.
If the QFI of a two-qubit state, with $(\hat{S}^{z}_{\textrm{e}}+\hat{S}^{z}_{\textrm{c}})$ being
the generator, is larger than 2, i.e., $\mathcal{Q}'(t;\phi_1)>2$, it characterises the useful
entanglement for quantum-enhanced parameter estimation \cite{Giovannetti2011}.
%Given
%$\hat{S}^{z}_{\textrm{e}}+\hat{S}^{z}_{\textrm{c}}=\hat{\sigma}^z_{\textrm{e}}/2+\hat{\sigma}^z_{\textrm{c}}/2$
%the generator, if QFI of the two-qubit system is larger than 2: $\mathcal{Q}'(t;\phi_1)>2$, it characterises the useful
%resource for the entanglement-enhanced parameter estimation \cite{Giovannetti2011}.

The time evolution of the QFI of the maximally entangled state is shown in Fig.~4a, when the controllable
 ${}^{14}$N channel is  either closed ($\phi_1=0$) or open ($\phi_1=\pi/2$). At time $t=0$ with $\phi_1=0$, we
 obtain the maximum QFI, $\mathcal{Q}'(0;0)=3.687$, which is useful for sub-shot-noise-limit
 metrology \cite{Giovannetti2011,Liu2015}.
  With the dissipative channel of  ${}^{14}$N closed ($\phi_1=0$), the QFI flow remains negative, except for
few time intervals due to the quantum fluctuations (see Fig.~4b).
 In Fig.~4c, the positive
 QFI flow of the maximally entangled state, with $\phi_1=\pi/2$, clearly signals the
 non-Markovian dynamics of the two-qubit open system.
 Moreover, the metrologically useful entanglement
  [$\mathcal{Q}'(t;\phi_1)>2$] survives for a period of time ($\lesssim380$~ns) with respect to the Markovian noise
 of the spin bath. However, it decays faster under the impact of non-Markovian noise by
 setting $\phi_1=\pi/2$.

Our experiments clearly demonstrate the engineering of the non-Markovian dynamics of
open systems by manipulating the electron spin, the  host $^{14}$N, and the neighbouring ${}^{13}$C nuclear spins
of the NV centre in diamond at room temperature. %The non-Markovianity is
%witnessed and quantified by QFI flows, which characterize the flow of quantum coherence and
%metrologically useful entanglement between the open systems and the environments.
First, the electron
qubit, as an open system, is subject to two controllable dissipative channels of nearby nuclear qubits, of
which the QFI flow  characterises the non-Markovianity and can be
decomposed into subflows from individual channels. Second, when the
open system, consisting of the electron qubit and the ${}^{13}$C qubit, is prepared in the maximally
entangled state, the controllable non-Markovian behaviour of the decoherence dynamics of the
entanglement witnessed by the QFI flow is reported.
By using the QFI as a witness for quantum coherence and metrologically useful entanglement,
our work will contribute to the developments of both noisy quantum metrology and
the non-Markovian dynamics of quantum open systems in solids \cite{Liu2015}.

%In details,

%In this Letter, we experimentally demonstrate and witness tunable quantum non-Markovian dynamics in a bulk diamond from a bran-new quantity quantum Fisher information (QFI) flow by regarding the electron spin as the system of interest and the nitrogen and strongly coupled 13C nuclear spin as decay channels. Moreover, the experimental results exactly show that the total QFI flow can be decomposed into split contributions from individual channel at any fixed evolution time. Additionally, we extend the demonstration in QFI flow from quantum coherence to quantum entanglement by taking the  electron spin coupled to a 13C nuclear spin as the whole system to measure quantum non-Markovianity. It is worthy pointing out that QFI characterizes highest precision to estimate an unknown parameter according to quantum Cramer-Rao theorem in quantum metrology.

%Quantum Speed Limits in Open System Dynamics through noisy QFI  \cite{Taddei2013}.
%Quantum metrology in solids .

%\emph{Introduction.---}%

%\emph{The toric code model.---}%
%
%
%
%
%
%%\emph{The dual transformation and dual quantum Fisher information.---}%
%\emph{The Wilson loop and dual transformations.---}%
%
%
%\emph{Thermalization of dual multipartite entanglement of topologically ordered states after a quantum quench.---}%
%
%
%\emph{Capabilities of a disorder protected topological order.---}%
%
%\emph{Conclusions.---}%

%\appendix
~

\noindent{\bf Acknowledgements}
We would like to acknowledge Zi-Yong Ge and Qing Ai for helpful discussions, and  Da Wei Lu for the codes of maximum likelihood estimation.
This work was supported by the National Key Research and Development
Program of China (Grant Nos.~2019YFA0308100,
2016YFA0302104, 2016YFA0300600), Strategic Priority Research
Program of Chinese Academy of Sciences (Grant
No.~XDB28000000), and NSFC (Grant Nos.~11974020,
11934018, 11574386). %, U1930402).
Y.R.Z. was supported by the China Postdoctoral Science
Foundation (Grant No.~2018M640055) and the  JSPS Postdoctoral Fellowship (Grant No.~P19326).
%J.Q.Y. was supported in part by: .
F.N. was supported in part by: the AFOSR (Grant No.~FA9550-14-1-0040), the ARO (Grant No.~W911NF-18-1-0358), the JST  Q-LEAP program, the JST CREST (Grant No.~JPMJCR1676), the JSPS-RFBR (Grant No.~17-52-50023, the JSPS-FWO (Grant No.~VS.059.18N), the RIKEN-AIST Challenge Research Fund, FQXi,
and the NTT PHI Lab.
%H.F. was partially supported by Ministry of Technology of China (grants No. 2016YFA0302104
%and 2016YFA0300600), National Natural Science Foundation of China  (grant No.~11774406) and Chinese Academy of Sciences (grants No.~XDPB-0803). J.Q.Y. was partially supported
%by the National Key Research and Development Program
%of China (grant No.~2016YFA0301200), the NSFC (grant No.~11774022),
%and the NSAF (grant No.~U1530401). F.N. was partially supported by the
%MURI Center for Dynamic Magneto-Optics via the AFOSR Award No.~FA9550-14-1-0040,
%the Japan Society for the Promotion of Science (KAKENHI), the IMPACT program of JST,
%CREST Grant No.~JPMJCR1676,  RIKEN-AIST Challenge Research Fund,
%JSPS-RFBR Grant No.~17-52-50023, and the Sir John Templeton Foundation.

~

\noindent{\bf Author Contributions}
G.Q.L., X.Y.P., and H.F. supervised
the project. Y.R.Z. and G.Q.L. conceived the idea. Y.R.Z.,
Y.N.L., and G.Q.L. designed the experimental schemes. Y.N.L
carried out the experimental measurements. Y.R.Z., Y.N.L., and
G.Q.L. analysed the data.  Y.R.Z and
Y.N.L. performed the numerical simulations. Y.R.Z., Y.N.L.,
G.Q.L., and F.N. wrote the manuscript.  All authors commented
on the manuscript.
%
%      Y.R.Z., Y.N.L., G.Q.L., and H.F. designed the experimental schemes. Y.N.L. conducted the experiment.  Y.R.Z., Y.N.L., and G.Q.L. analysed the data. Y.R.Z. and Y.N.L. performed the numerical simulations. X.Y.P. built the experimental platform.
%     All authors discussed the results and contributed to the
%     manuscript.

     ~

\noindent{\bf Competing Interests}
  The authors declare no competing interests.

~
%\clearpage

\noindent\textbf{METHODS}

~

\noindent\textbf{QFI flow and non-Markovianity.}
Consider a mixed state $\rho=\sum_j\lambda_j|j\rangle\langle j|$, with $\langle i|j\rangle=\delta_{ij}$ and a generator
$\hat{\mathcal{O}}$, the QFI of  a state $\rho(\theta)=\exp(-i\theta\hat{\mathcal{O}})\rho\exp(-i\theta\hat{\mathcal{O}})$
with respect to a parameter $\theta$ can be written as  \cite{BRAUNSTEIN1994}
\begin{equation}
\mathcal{Q}=2\!\!\sum_{\lambda_i+\lambda_j\neq0}\!\!\frac{(\lambda_i-\lambda_j)^2}{\lambda_i+\lambda_j}|\langle i|\hat{\mathcal{O}}|j\rangle|^2.\nonumber
\end{equation}
For a single qubit,  any state can be expressed as
\begin{equation}
\rho={\mathbb{I}/2+\!\!\sum_{\alpha=x,y,z}\!\!s_\alpha\,\hat{\sigma}_\alpha}/{2},\nonumber
%=\frac{1}{2}\left[\begin{array}{cc}1+s_z&s_
\end{equation}
with $\bm{r}=[s_x,~s_y,~s_z]$ being the Bloch vector and $r\equiv({s_x^2+s_y^2+s_z^2})^{\frac{1}{2}}$ being the Bloch length.
The QFI with respect to the generator $\hat{\mathcal{O}}=\hat{\sigma}_z/2$ can be calculated as
$\mathcal{Q}=r^2-s_z^2$.
For a qubit state with $\bm{s}=[1,~0,~0]$,  its QFI can be calculated to be 1.
The QFI plays a central role in quantum metrology and multipartite entanglement witness \cite{Toth2012,Pezze2009,Zhang2018}, which is also
sufficient for measuring non-Markovianity \cite{Lu2010}.

We then consider a quantum process, described by a time-local master equation \cite{Breuer2016,Rivas2014}
\begin{equation}
\frac{\partial\rho}{\partial t}=-i[\hat{H},\rho]+\sum_j{\gamma_j}\left(\hat{A}_j\rho \hat{A}_j^\dag-{\{\hat{A}_j^\dag \hat{A}_j,\rho\}}/{2}\right),\nonumber
\end{equation}
where $\hat{H}$ is the Hamiltonian for the open system without coupling to the bath, $\gamma_j(t)$
is the time-dependent decay rate, and $\hat{A}_j(t)$ is the time-dependent Lindblad operator.
The QFI flow of the quantum open system can be divided into different sub-channels as \cite{Lu2010}
$\mathcal{I}\equiv {\partial \mathcal{Q}}/{\partial t}=\sum_j\mathcal{I}_j$, %=\sum_j\gamma_j(t)\mathcal{J}_j
with $\mathcal{I}_j=\gamma_j(t)\mathcal{J}_j$, and $\mathcal{J}_j\leq0$. Therefore, the existence of any
positive QFI subflow characterises a quantum non-Markovian process based on the completely positive divisibility \cite{Breuer2016}.
%the CPdivisible
%dynamics

%\section{Experimental details}
%In this section, we mainly describe the measurement setup, sample information and state tomography techniques in detail.
~

\noindent\textbf{Measure of non-Markovianity via QFI subflows.}
From Eq.~(\ref{measure}), the non-Markovianity can be witnessed by the sum of time integrals of all inward QFI subflows.
When $\phi_1=\phi_2=\pi/2$, QFI subflows $\mathcal{I}_{\textrm{n}}(t)$, $\mathcal{I}_{\textrm{c}}(t)$ and
$\mathcal{I}_{\textrm{R}}(t)$ can be calculated from QFI
with different parameters: $\mathcal{Q}_{\textrm{R}}\equiv\mathcal{Q}(t,0,0)$, $\mathcal{Q}_{\textrm{nR}}\equiv\mathcal{Q}(t,0,\frac{\pi}{2})$, and $\mathcal{Q}_{\textrm{cR}}\equiv\mathcal{Q}(t,0,\frac{\pi}{2})$, which
can be expressed, with $\dot{\mathcal{Q}}\equiv{\partial \mathcal{Q}}/{\partial t}$, as
\begin {align}
\mathcal{I}_{\textrm{n}}&=(\dot{\mathcal{Q}}_{\textrm{nR}}-{\mathcal{Q}_{\textrm{nR}}\dot{\mathcal{Q}}_{\textrm{R}}}/{\mathcal{Q}_{\textrm{R}}}){\mathcal{Q}_{\textrm{cR}}}/{\mathcal{Q}_{\textrm{R}}},\nonumber\\
\mathcal{I}_{\textrm{c}}&
=(\dot{\mathcal{Q}}_{\textrm{cR}}-{\mathcal{Q}_{\textrm{cR}}\dot{\mathcal{Q}}_{\textrm{R}}}/{\mathcal{Q}_{\textrm{R}}}){\mathcal{Q}_{\textrm{nR}}}/{\mathcal{Q}_{\textrm{R}}},\nonumber\\
\mathcal{I}_{\textrm{R}}&={{\mathcal{Q}}_{\textrm{cR}}\mathcal{Q}_{\textrm{nR}}\dot{\mathcal{Q}}_{\textrm{R}}}/
{\mathcal{Q}_{\textrm{R}}^2}.\nonumber
\end{align}

~

\noindent\textbf{Setup and sample.}
Our experiments are performed under ambient conditions on a high-purity bulk diamond (Element Six, with N concentration $< 5$~p.p.b., and natural abundance of $^{13}$C isotopes).
The NV centre is located 10~$\mu $m  below the diamond surface.
To enhance the photon collection efficiency of the NV centre, solid immersion lenses (SILs) are etched on the diamond surface \cite{Marseglia2011}.
The photon detection rate is 450~kcps, when the  laser (532~nm) excitation power
is  240~$\mu$W.

We use the ODMR to detect nuclear spins, which are
strongly coupled to the NV electron spin. From the continuous-wave ODMR spectrum under
a small magnetic field $B\simeq$40~Gauss  (see Extended Data Figs.~1a and 1b),
the host $^{14}$N nuclear spin and the nearby $^{13}$C nuclear spin with 12.8~MHz coupling
strength are identified \cite{Smeltzer2011}.
 To identify the nuclear spin with a weaker coupling strength, the external magnetic field is tuned to
 482~Gauss (along the quantisation axis of the NV centre), and both the host $^{14}$N and
 the 12.8~MHz $^{13}$C nuclear spins can be polarised by a short laser pulse.
 We then measure the pulsed-ODMR spectra of the NV centre (see Extended Data Fig.~1c).
By reducing the MW power,  two other $^{13}$C nuclear spins are resolved (see Extended Data Fig.~1d).
The hyperfine splittings caused by these two $^{13}$C nuclear spins are 0.9~MHz and 0.4~MHz, respectively.
 These two nuclear spins are partially polarised by the laser pumping (see Extended Data Fig.~1d),
 which is consistent with the literature results \cite{Dreau2012}.
In our experiments,  the 12.8 MHz $^{13}$C nuclear spin and the $^{14}$N nuclear
 spin are taken as fully controllable decoherence channels, while the other nearby weakly-coupled $^{13}$C nuclear spins behave as uncontrollable
 decoherence channels.

 ~

 \noindent\textbf{Coherent manipulation of the NV centre.}
%Extended data Fig.~1e and 1f shows the spin coherence of the NV centre.
The electron spin and nuclear spins are manipulated by the resonant MW and RF pulses.
Extended Data Fig.~1e shows a typical Rabi oscillation signal of the NV electron spin, with Rabi frequency 23.8~MHz.
Extended Data Fig.~1f shows the Ramsey oscillation from a deliberate MW detuning (2~MHz) and the beating between different transitions, according to the state of the 0.4~MHz $^{13}$C nuclear spin.
By fitting the Ramsey signal, we obtain  the coherence time of this NV centre as $T^{\star}_{2}\approx 2.9~\mu$s.

%The electron spin and nuclear spins are controlled by the microwave and radio-frequency currents,
%which are delivered through a fabricated coplanar waveguide on a PCB board. In Extended data Fig.~1e,
%the Rabi frequency of the electron spin is about 23.8~MHz.  In Extended data Fig.~1f, the oscillation of the
%Ramsey signal is obtained from the deliberate microwave detuning and the beating between different
%transition lines, according to the state of the 0.4~MHz $^{13}$C nuclear spin. The coherence time of the NV
%centre is $T^{\star}_{2}\approx 2.9~\mu s$.

~

\noindent\textbf{State tomography.}
%The density matrix of any single qubit can be written as
%\begin{equation}
%\rho=\frac{1}{2}\left( \begin{array}{cc}
%S_{0}+S_{3} & S_{1}-iS_{2}\\
%S_{1}+iS_{2}& S_{0}-S_{3}
%\end{array}\right)
%\label{matrix}
%\end{equation}
%where $S_{0}=1$ due to $Tr(\rho)=1$.\\
%
%
%\indent
The sequence to realise the single-qubit state tomography can be divided into three parts:
($i$) Directly collect fluorescence photon
counts $L_z$; ($ii$) prepare the state, apply a $\frac{\pi}{2}$ MW pulse along the $x$-axis, and then collect the
fluorescence photon counts $L_y$;  ($iii$)
prepare the state,
apply a $\frac{\pi}{2}$ MW pulse along the $y$-axis, and then collect the
fluorescence photon counts $L_x$.
In addition, we collect the fluorescence photon counts $L_{0}$ in the bright state ($ m_{s}=0$),
and $L_{1}$ in the dark state ($ m_{s}=-1$), as references to normalise the florescence signal.
With the signals on different bases, we can reconstruct the state by calculating
the Bloch vector $\bm{r}=[s_x,~s_y,~s_z]$ with
\begin{align}
s_x&=-(2L_x-L_0-L_1)/(L_0-L_1),\label{eq18}\\
s_y&=(2L_y-L_0-L_1)/(L_0-L_1),\label{eq19}\\
s_z&=(2L_z-L_0-L_1)/(L_0-L_1).
\label{eq20}
\end{align}

The two-qubit state tomography technique is based on the single-qubit state tomography method.
The key point is to apply one or two additional transfer pulses (RF/MW pulses, see concrete sequences in
 Ref.~\cite{Liu2013}) to divide the
full 4$\times$4 density matrix of two qubits into several 2$\times$2 reduced density matrices.
Then, we can in turn obtain the real and imaginary parts of each element of the density matrix by implementing the single-qubit state
tomography in each working transition (see Refs.~\cite{Liu2013,Van2012}).

~

\noindent\textbf{QFI envelope induced by the spin bath.}
To quantitatively describe the influence of the spin bath on the dynamics of the quantum open system, we measure
the long-term QFI of the NV electron spin, when all the controllable channels (12.8~MHz $^{13}$C and 2.16~MHz $^{14}$N)
are turned off after polarising by a short laser (see Extended Data Fig.~2b).
In Extended Data Figs.~2a and 2b, we find that the coherence and QFI decay non-monotonously, with a 0.4~MHz oscillation.
We attribute this oscillation to the effects of the partially polarised $^{13}$C nuclear spin, of which the hyperfine coupling strength is 0.4~MHz, and the quantisation axis is not the same as the NV electron spin's \cite{Dreau2012}.
Therefore, we can only regard the 0.4~MHz  $^{13}$C nuclear spin  as an uncontrollable quantum channel
 and fit the experimental results using the formula
\begin{equation}
\mathcal{Q}_{\textrm{R}}(t)=\exp [-(t/T_2^*)^\alpha] [1-\sin^2\phi_0\sin^2({A^{\|}_{c0}t}/2+\varphi_0/2) ],\nonumber
\end{equation}
with $\alpha=0.89$, $T_{2}^*\simeq 1026$~ns, $\phi_0=0.37\pi$, $A^{\|}_{c0}=0.4$~MHz, and $\varphi_0=0.21\pi$.
However, within the time scale of our experiments (0--600~ns), the dissipative channel of the 0.4~MHz  $^{13}$C nuclear spin behaves Markovian, which can be included in the Markovian channel of the
spin bath (see Extended Data Fig.~2b).
%In the main text, the QFI from spin bath dissipative channel is described by this formula with and the same parameters.

~

\noindent\textbf{Data processing.}
The measurements in our  experiments are repeated at least 4$\times$10$^{5}$ times to obtain a good signal-to-noise ratio.
In the first experiment, each data point of the QFI is calculated from the measured photon counts
($L_{x}$, $L_{y}$, $L_{z}$, $L_{0}$, and $L_{1}$) using Eqs.~(\ref{eq18}--\ref{eq20}).  After calculating the QFI, we smooth the data by averaging 5 data points around each specific point (the adjacent average smoothing method)
to calculate the QFI flow.
In the second experiment, each data point of the QFI is obtained from the full 4$\times$4 density matrix of two qubits. The final density matrix is extracted from the measured photon counts, according to the two-qubit state tomography technique, and optimised by a maximum likelihood estimation (MLE) to reduce measurement errors.

\bibliography{manuscript}

%\clearpage

%\clearpage

\renewcommand\thefigure{Extended Data \arabic{figure}}
\setcounter{figure}{0}

\begin{figure*}[t]
	\includegraphics[width=0.99\textwidth]{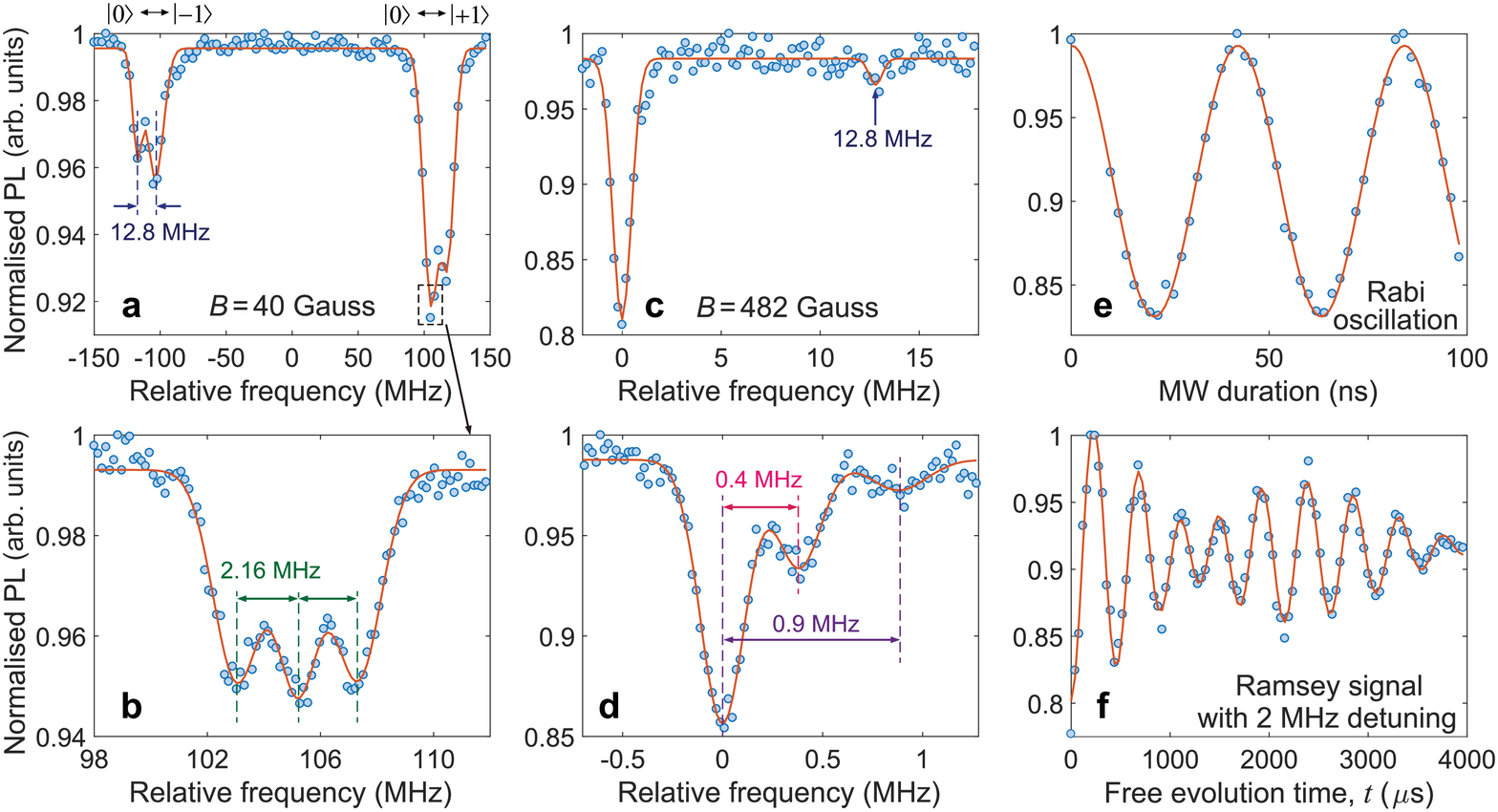}	
	\caption{{\bf Characterisation of the multi-qubit system.}
		(\textbf{a},\textbf{b}) Continuous-wave ODMR spectra under a small magnetic field with
		$B=40$~Gauss (frequency is relative to 2,870~MHz).
		The 2.16~MHz and 12.8~MHz splittings are induced by the $^{14}$N nuclear spin and
		 the nearby $^{13}$C nuclear spin, respectively.
		(\textbf{c},\textbf{d}) Pulsed-ODMR spectra under a magnetic field with $B = 482$~Gauss
		(frequency is relative to 1,518~MHz).
		Both the host $^{14}$N nuclear spin and the 12.8 MHz $^{13}$C nuclear spin are polarised due to ESLAC,
		and two other nuclear spins (0.9~MHz and 0.4~MHz) can be resolved.
		%shows that the nearby $^{13}$C and $^{14}$N nuclear spins are both polarised by laser pumping due to ESLAC.
		%(\textbf{d}) The pulse-ODMR spectrum verifies the existence of another nearby $^{13}$C nuclear spins by reducing the MW power.
		(\textbf{e}) The Rabi oscillation and (\textbf{f}) Free induction decay of the NV electron spin (with 2 MHz detuning).
		The coherence time of the NV electron spin is $T^*_2\simeq 2.9$~$\mu$s.
	}
	\label{figS1}
\end{figure*}

\begin{figure*}[b]
	\includegraphics[width=0.72\textwidth]{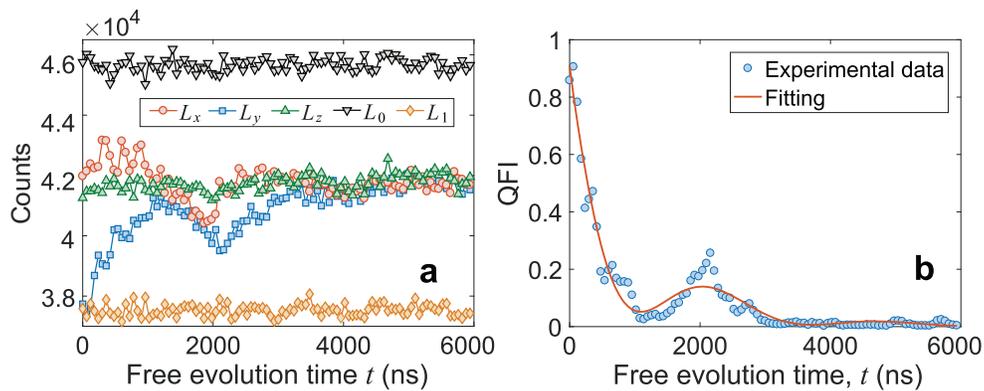}	
	\caption{{\bf Long-time QFI of the electron spin coupled to the uncontrolled spin bath.}
		(\textbf{a}) Measured fluorescence photon counts as functions of the free evolution time.
		(\textbf{b})  QFI of the electron qubit as a function of the evolution time, where the oscillation is induced by the nearby partially polarised 0.4~MHz $^{13}$C nuclear spin.
	}
	\label{figS3}
\end{figure*}

\end{document}